\documentclass{article}\usepackage[]{graphicx}\usepackage[]{color}

\usepackage[numbers]{natbib}
\usepackage{color}
\usepackage{tikz}
\usepackage{amsmath}
\usepackage{url}
\usepackage[left=2.5cm,right=2.5cm, top = 2.5cm, bottom = 3cm]{geometry}

\begin{document}

\title{Comment on ``Under-reported data analysis with INAR-hidden Markov chains''}

\author{Johannes Bracher}

\maketitle

\begin{center}
Epidemiology, Biostatistics and Prevention Institute, University of Zurich,\\ Hirschengraben 84, 8001 Zurich, Switzerland\\

\medskip

\texttt{johannes.bracher@uzh.ch}

\bigskip

\noindent\fbox{\parbox{0.8\textwidth}{
\textbf{Bibliographic note:} This is the pre-reviewing version of a letter published in \textit{Statistics in Medicine} 38(5), 893--898:\\
\url{https://onlinelibrary.wiley.com/doi/full/10.1002/sim.8032}\\
Fern\'andez-Fontelo et al published a reply which raises some interesting further points:\\
\url{https://onlinelibrary.wiley.com/doi/10.1002/sim.8033}\\
After a 12 month embargo period the accepted version will be made available on \textit{arxiv}.
}}

\end{center}

\bigskip

\abstract{In Fern\'andez-Fontelo et al (Statis. Med. 2016, DOI 10.1002/sim.7026) hidden integer-valued autoregressive (INAR) processes are used to estimate reporting probabilities for various diseases. In this comment it is demonstrated that the Poisson INAR(1) model with time-homogeneous underreporting can be expressed equivalently as a completely observed INAR($\infty$) model with a geometric lag structure. This implies that estimated reporting probabilities depend on the assumed lag structure of the latent process.}

\section{Introduction}
\label{sec:introduction}

We read with great interest the article by Fern\'andez-Fontelo et al \cite{Fernandez-Fontelo2016} who discuss underreporting in count time series models, a problem which is encountered in many real-world settings. In case studies, reporting probabilities for several diseases are estimated under the assumption of a hidden INAR(1) process. In our comment we develop this idea further and point out an identifiability problem which arises when the INAR(1) assumption is relaxed. Specifically it is shown that it is not possible to distinguish between time-homogeneous underreporting and a geometric lag structure in Poisson INAR models. Estimation of reporting probabilities from time series data thus relies on the correct specification of the lag structure of the latent process.

The Poisson INAR(1) model \cite{Al-Osh1987} with parameters $\lambda > 0$ and $0 \leq \alpha < 1$ is defined as
\begin{equation}
X_t = \alpha \circ X_{t - 1} + W_t \label{eq:inar1}
\end{equation}
where the $W_t$ independently follow a $\text{Poisson}(\lambda)$ distribution. The operator $\circ$ denotes binomial thinning, i.e.\ $\alpha \circ X_{t - 1} = \sum_{i = 1}^{X_{t - 1}} Z_i$ with $Z_i \stackrel{\text{iid}}{\sim} \text{Bern}(\alpha)$. The thinning operations are assumed to be independent of each other and of $\{W_t\}$; further, the thinning operations at each $t$ and $W_t$ are assumed to be independent of $X_{t - 1}, X_{t - 2}, \dots$. Fern\'andez-Fontelo et al \cite{Fernandez-Fontelo2016} introduce an underreported version of this model. However, they assume that counts are only subject to underreporting with a certain probability $\omega$; otherwise reporting is assumed to be complete. The observed process is thus
\begin{equation}
\tilde{X}_t = \begin{cases}
X_t & \text{ with probability } 1 - \omega\\
q \circ X_t & \text{ with probability } \omega,
\end{cases}\label{eq:reporting_fernandez}
\end{equation}
where $0 < q \leq 1$ is a reporting probability and, given $X_t$, $\tilde{X}_t$ is independent of the past. The assumption that reporting is 100\% complete during some periods seems strong in many contexts. And indeed, in two case studies (weekly number of human papillomavirus cases in Girona, Spain; annual deaths from mesothelioma in Great Britain) the estimates $\hat{\omega}$ are close to 1 with confidence intervals including this value. This indicates that $\omega = 1$, i.e.\ \textit{time-homogeneous underreporting}, is an important special case. In the following the focus will thus be on models where, instead of \eqref{eq:reporting_fernandez}, the reporting process is
\begin{equation}
\tilde{X}_t = q \circ X_t\label{eq:reporting_simple}\ .
\end{equation}

\section{Interplay of underreporting and geometric lags in INAR models}
\label{subsec:equivalence}

We now generalize the INAR(1) from \eqref{eq:inar1} to an INAR($\infty$) model with geometric lags and show how closely this aspect is related to underreporting. The starting point is the INAR($p$) model introduced by Alzaid and Al-Osh \cite{Alzaid1990} which is defined as
\begin{align}
X_t & = \sum_{i = 1}^p \alpha_i \circ X_{t - i} \ \ + \ \ W_t\label{eq:X}\\
(\alpha_1 \circ X_t, \dots, \alpha_p \circ X_t) \mid X_t & \sim \text{Mult}(\alpha_1, \dots, \alpha_p, X_t)\label{eq:multinomial}
\end{align}
with $\sum_{i = 1}^p \alpha_i < 1$.

This INAR($p$) model does not allow for $p = \infty$ as the multinomial distribution requires a finite number of categories. We therefore reformulate the multinomial distribution in \eqref{eq:multinomial} as (compare e.g.\ \cite{Johnson1997}, p. 33)
\begin{align}
B_t \ \mid  \ X_t & \sim \text{Bin}\left(X_t, \sum_{i = 1}^p \alpha_i\right)\\
A^{(j)}_t & = \begin{cases} 1 & \text{ with probability } \alpha_1/\sum_{k = 1}^p \alpha_k \\
                    \vdots\\
                    p & \text{ with probability } \alpha_p/\sum_{k = 1}^p \alpha_k \end{cases} \ \ \ \ \ \text{ i.i.d. for } j = 1, \dots, B_t \label{eq:waiting_timesv2}\\
\alpha_i \circ X_{t} & =\sum_{j = 1}^{B_{t}} I(A^{(j)}_{t} = i), \ \ \ i = 1, \dots, p \label{eq:coupling_thinning_p}
\end{align}
where $I$ is the indicator function. In the interpretation given in Wei{\ss} \cite{Weiss2018} (p. 46), $B_t$ is the number of individuals (among the $X_t$ from time $t$) which will be \textit{renewed} during $t + 1, \dots, t + p$. The variables $A_t^{(j)}, j = 1, \dots, B_t$ are the respective waiting times. This formulation extends more easily to the case $p = \infty$, specifically consider a geometric lag structure
\begin{equation}
\alpha_i = \beta \gamma^{i - 	1}, i = 1, 2, \dots\label{eq:alpha_inar}
\end{equation}
with $0 < \beta < 1 - \gamma < 1$. As $\sum_{i = 1}^\infty \alpha_i = \beta/(1 - \gamma)$ we get $\alpha_i /\left( \sum_{k = 1}^\infty \alpha_k \right)= (1 - \gamma)\gamma^{i - 1}$, i.e.\ the $A_t^{(j)}$ follow a geometric distribution with parameter $1 - \gamma$ and support $\{1, 2, \dots\}$. The INAR($\infty$) process $\{X_t\}$ with parameters $\beta$ and $\gamma$ can thus be defined as
\begin{align}
X_t & = \sum_{i = 1}^\infty \alpha_i \circ X_{t - i} \ \ + \ \ W_t \label{eq:inar_inf}\\
B_t \ \mid X_t & \sim \text{Bin}(X_t, \beta/(1 - \gamma))\\
A_{t}^{(j)} & \stackrel{\text{iid}}{\sim} \text{Geom}(1 - \gamma), \ \ \ j = 1, \dots B_{t}\label{eq:At}\\
\alpha_i \circ X_{t} & =\sum_{j = 1}^{B_{t}} I(A^{(j)}_{t} = i), \ \ \ i = 1, 2,\dots \label{eq:coupling_thinning_geom}
\end{align}
This ensures, like the multinomial distribution in \eqref{eq:multinomial}, that $(\alpha_i \circ X_t) \mid X_t \sim \text{Bin}(X_t, \alpha_i)$ while $\sum_{i = 1}^\infty \alpha_i \circ X_t \leq X_t$. Again for each $t$ it is assumed that $W_{t}$ and $\alpha_i \circ X_t \mid X_t, i = 1, \dots$ are independent of the past history, i.e. $X_{t - k}$ and $\alpha_j \circ X_{t - k}$ for $j, k = 1, 2, \dots$. Such geometric lag structures are closely linked to underreporting as the following can be shown (proofs in the Appendix):
\begin{enumerate}
\item[(A)] Consider a Poisson INAR(1) process $\{X_t\}$ with parameters $\lambda_X, \alpha_X$ and the underreported process $\{\tilde{X}_t = q_X \circ X_t\}$. There is a Poisson INAR($\infty$) process $\{Y_t\}$ of type \eqref{eq:inar_inf}--\eqref{eq:coupling_thinning_geom} which is is equivalent to $\{\tilde{X}_t\}$; its parameters are $\lambda_Y = \lambda_X q_X/\{1 - \alpha_X(1 - q_X)\}; \beta_Y = \alpha_X q_X; \gamma_Y = \alpha_X(1 - q_X)$.

\item[(B)] Consider a Poisson INAR($\infty$) process $\{X_t\}$ of type \eqref{eq:inar_inf}--\eqref{eq:coupling_thinning_geom} with parameters $\lambda_X, \beta_X$ and $\gamma_X$. The underreported process $\{\tilde{Y}_t = q_Y \circ Y_t\}, q_Y = \beta_X/(\beta_X + \gamma_X)$ where $Y_t$ is an INAR(1) process with parameters
$\lambda_Y = \lambda_X(\beta_X + \gamma_X)(1 - \gamma_X)/\beta_X; \alpha_Y = \beta_X+\gamma_X$
is equivalent to $\{X_t\}$.
\item[(C)] More generally, consider a Poisson INAR($\infty$) process $\{X_t\}$ with parameters $\lambda_X, \beta_X, \gamma_X$ and $\{\tilde{X}_t = q_X \circ X_t\}$. For any $q_Y \in [q_X\beta_X/(\beta_X + \gamma_X), 1]$ there is a Poisson INAR($\infty$) process $\{Y_t\}$ so that $\{\tilde{Y}_t = q_Y \circ Y_t\}$ is equivalent to $\{\tilde{X}_t\}$. The parameters of $\{Y_t\}$ are 
$$
\lambda_Y = \frac{\lambda_X (1 - \gamma_X) \frac{q_X}{q_Y}}{1 - \gamma_X - \left(1 - \frac{q_X}{q_Y}\right)\beta_X}\ ; \ \ \ \beta_Y = \beta_X \frac{q_X}{q_Y}\ ; \ \ \ \gamma_Y = \gamma_X + \left(1 - \frac{q_X}{q_Y}\right)\beta_X\ .
$$
\end{enumerate}
For each underreported INAR($\infty$) process and, as a special case, each underreported INAR(1) process, there are thus many different INAR($\infty$) representations. Each of them features a different combination of reporting probability $q_X$ and decay factor $\gamma_X$ for the autoregressive parameters. In Fern\'andez-Fontelo et al \cite{Fernandez-Fontelo2016} identifiability is ensured by the assumption that the latent process is indeed INAR(1), i.e.\ $\gamma_X = 0$. Without substantial prior knowledge to justify a precise value of $\gamma_X$, however, the reporting probability $q_X$ cannot be estimated.

\section{Application to human papillomavirus cases in Girona, Spain}

\begin{figure}[h!]
\center
\includegraphics{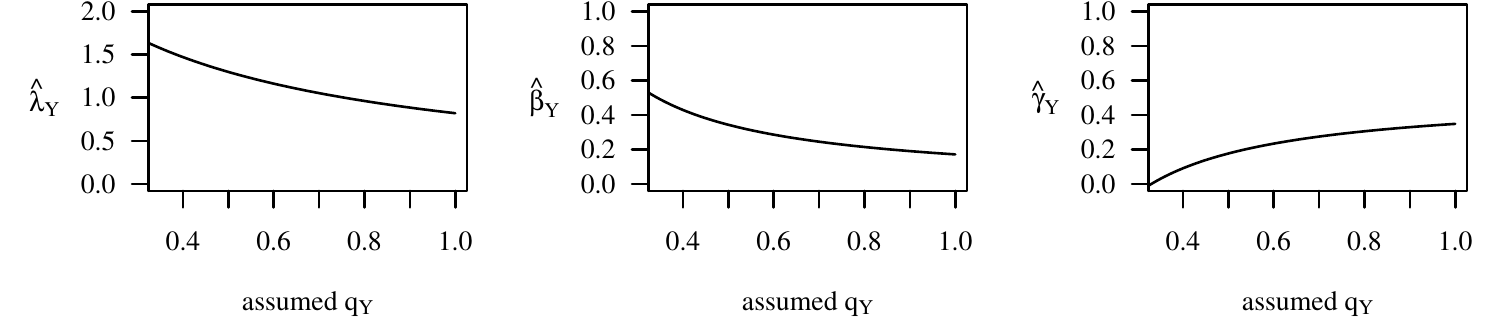}
\caption{Parameter values of underreported INAR($\infty$) models with different reporting probabilities which are all equivalent to an underreported INAR(1) model with parameters $\hat{\alpha}_X = 0.52, \hat{\lambda}_X = 1.62, \hat{q}_X = 0.33$ (eq. \eqref{eq:estimated_mod_X}--\eqref{eq:estimated_mod_X_tilde}).}
\label{fig:q_star}
\end{figure}

For illustration of the argument we revisit the analysis of reported human papillomavirus cases in Girona, Spain (2010--2014) from Fern\'andez-Fontelo et al \cite{Fernandez-Fontelo2016}. The authors assume the latent model \eqref{eq:inar1} with reporting process \eqref{eq:reporting_fernandez}, but as $\hat{\omega} = 0.92$ with a confidence interval from 0.78 to 1.07 they suggest that the simpler version \eqref{eq:reporting_simple} could be used as well. We will thus simplifyingly pretend that their parameter estimates come from this simpler model (even though the model then cannot accomodate overdispersion; this could be addressed via a different immigration distribution). The estimated data generating process (eq. (22)--(23) in \cite{Fernandez-Fontelo2016}) is then
\begin{align}
X_t & = 0.52 \circ X_{t - 1} + W_t(1.62)\label{eq:estimated_mod_X}\\
\tilde{X}_t & = 0.33 \circ X_t\ .\label{eq:estimated_mod_X_tilde}
\end{align}
The most interesting result from a public health perspective is the estimated reporting probability of 0.33 as it directly translates to an estimate of the unobserved disease burden. Using statement (C) from Section \ref{subsec:equivalence}, however, we can pick any $q_Y \in [0.33, 1]$ and obtain the parameters $\hat{\beta}_Y, \hat{\gamma}_Y, \hat{\lambda}_Y$ of an INAR($\infty$) process $\{Y_t\}$ so that $\{\tilde{Y}_t = q_Y \circ Y_t\}$ is equivalent to $\{\tilde{X}_t\}$. In Figure \ref{fig:q_star}these parameters are diplayed as functions of $q_Y$.

So if we relax the INAR(1) assumption from the original article and assume a latent INAR($\infty$) process, the data are equally compatible with a whole range of reporting probabilities. The most extreme re-formulation of $\{\tilde{X}_t\}$ from equations \eqref{eq:estimated_mod_X}, \eqref{eq:estimated_mod_X_tilde} is a completely observed INAR($\infty$) process $\{Y_t\}$ with
$$
Y_t = 0.17\circ Y_{t - 1} + 0.06\circ Y_{t - 2} + 0.02\circ Y_{t - 3}+ 0.01\circ Y_{t - 4} + \dots + W_t(0.82)
$$
where the lagged terms after $p = 4$ are cut off as the respective parameters become very small. In certain settings generation time distributions of infectious diseases may give us an idea about an appropriate lag structure and thus value of $\gamma_Y$, making the model identifiable. The period of communicability of HPV is unknown, but likely to be at least as long as the persistence of lesions \cite{Heymann2015}. Development of lesions is assumed to take 2--3 months in most cases. For weekly data a more spread-out lag structure may thus be more appropriate than an INAR(1) specification.

\section*{Appendix: Derivation of statements (A), (B), (C) from Section 2}

\begin{figure}[h]
\centering
\includegraphics{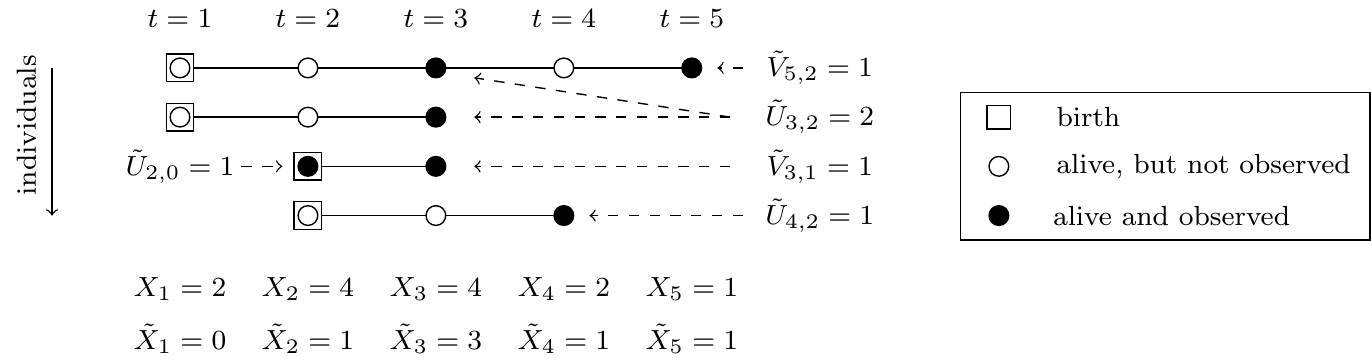}
\caption{Stylized example to illustrate notations \eqref{eq:U} and \eqref{eq:V}: individual-level display of a population of individuals following principles (i)--(iii). The total population size $\{X_t\}$ follows an INAR(1) process, but only an underreported version $\{\tilde{X}_t = q_X \circ X_t\}$ is observed. Non-zero values of the newly introduced auxiliary variables $\tilde{U}_{t, i}$ and $\tilde{V}_{t, i}$ are added in the graph.}
\label{fig:notation}
\end{figure}

Consider a Poisson INAR(1) process $\{X_t\}$ with
$$
X_t = \alpha\circ X_{t - 1} + W_{t}\ ; \ \ W_t \stackrel{\text{iid}}{\sim} \text{Pois}(\lambda_X)
$$
and the underreported process $\{\tilde{X}_t = q_X \circ X_t\}$. It is now shown that, as stated in (A), $\{\tilde{X}_t\}$ is equivalent to an INAR($\infty$) process $\{Y_t\}$ of type \eqref{eq:inar_inf}--\eqref{eq:coupling_thinning_geom} with parameters $\lambda_Y, \beta_Y, \gamma_Y$. The argument is easiest understood when expressed in terms of the survival interpretation of $\{\tilde{X}_t\}$:
\begin{itemize}
\item[(i)] New individuals can be born at each time step $t$; their number follows a Poisson distribution with rate $\lambda_X$.
\item[(ii)] Individuals already present at $t$ have a probability $\alpha_X$ of still being alive at time $t + 1$.
\item[(iii)] Alive individuals are observed with probability $q_X$ at each time step.
\end{itemize}
All births, deaths and observation events are assumed to be independent. $X_t$ is the number of individuals alive at time $t$, $\tilde{X}_t$ is the number of those who are observed. Now denote by $\tilde{U}_{t\cdot}$ the number of individuals observed in $t$ which \textit{have not been observed previously}, and by $\tilde{V}_{t\cdot}$ the number of individuals observed in $t$ which \textit{have already been observed} at a previous time point so that
\begin{equation}
\tilde{X}_t = \tilde{U}_{t\cdot} + \tilde{V}_{t\cdot} \ .\label{eq:X_tilde}
\end{equation}
The term $\tilde{U}_{t\cdot}$ can be further decomposed by when the individuals were born; denoting by $\tilde{U}_{t, i}, i = 0, 1, \dots$ the number of individuals first observed in $t$ and born in $t - i$ one gets
\begin{equation}
\tilde{U}_{t\cdot} = \sum_{i = 0}^\infty \tilde{U}_{t, i} \ .\label{eq:U}
\end{equation}
Similarly, $\tilde{V}_{t\cdot}$ is decomposed by when the individuals were last observed; denoting by $\tilde{V}_{t, i}, i = 1, 2, \dots$ the individuals last observed in $t - i$ and observed again in $t$ this leads to
\begin{equation}
\tilde{V}_{t\cdot} = \sum_{i = 1}^\infty \tilde{V}_{t, i} \ .\label{eq:V}
\end{equation}
Figure \ref{fig:notation} illustrates the definition of $\tilde{U}_{t, i}$ and $\tilde{V}_{t, i}$ with a simple example.

\noindent Obviously an individual born in $t$ can only be observed \textit{for the first time} in at most one out of $\{t, t + 1, \dots\}$, i.e.\ $W_t$ is split up into $U_{t, 0}, U_{t + 1, 1}, \dots$ and a part of individuals which is never observed. The probability that an individual born in $t$ is first observed in $t + i, i = 0, 1, \dots$ is $\alpha_X^{i}(1 - q_X)^{i}q_X$ (the individual has to survive $i$ times, stay unobserved $i$ times and finally be observed once). The splitting property of the Poisson distribution (\cite{Kingman1993}, 53) and the independence between the $W_t, t = 0, \pm 1, \dots$ then imply that all $\tilde{U}_{t, i}, i = 0, 1, \dots; t = 0, \pm 1, \dots$ independently follow Poisson distributions. Their rates depend on $i$, specifically $\tilde{U}_{t, i} \sim \text{Poisson}(\alpha_X^{i}(1 - q_X)^{i}q_X\lambda_X)$. Consequently, the $\tilde{U}_{t, \cdot}$, too, are independent of each other. As sums of independent Poisson random variables they likewise follow a Poisson distribution (\cite{Kingman1993}, 14):
\begin{equation}
\tilde{U}_{t, \cdot} \stackrel{\text{iid}}{\sim} \text{Poisson}(\lambda_Y); \ \ \lambda_Y = \sum_{i = 0}^\infty \{\alpha_X (1 - q_X)\}^i q_X\lambda_X = \frac{q_X\lambda_X}{1 - \alpha_X(1 - q_X)}\ .\label{eq:U_tilde}
\end{equation}
Similarly, an individual observed in $t$ can only be observed \textit{next} in at most one out of $\{t + 1, t + 2, \dots\}$. The $\tilde{X}_t$ observed individuals are thus split up into $\tilde{V}_{t + 1, 1}, \tilde{V}_{t + 2, 2}, \dots$ and a part of individuals which is never observed again. The probability that an individual is observed next in $t + i, i = 1, \dots$ is $\alpha_X^i(1 - q_X)^{i - 1}q_X$ (the individual has to survive $i$ times, stay unobserved $i - 1$ times and finally be observed once) or $\beta_Y\gamma_Y^{i - 1}$ with
\begin{equation}
\beta_Y = \alpha_X q_X, \ \ \gamma_Y = \alpha_X (1 - q_X).
\end{equation}
The probability that the individual will be observed again \textit{at all} is $\sum_{i = 1}^\infty \beta_Y\gamma_Y^{i - 1} = \beta_Y/(1 - \gamma_Y)$. Consequently, under the condition that the $j$-th of the $X_t$ individuals \textit{will} be observed again, the waiting time $A_t^{(j)}$ until this occurs has probability mass function $\text{Prob}(A_t^{(j)} = i) = \beta_Y\gamma_Y^{i - 1}/\{\beta_Y/(1 - \gamma_Y)\} = \gamma_Y^{i - 1}(1 - \gamma_Y), i = 1, \dots$. It thus follows a geometric distribution with parameter $1 - \gamma_Y$. Denoting the number of individuals observed in $t$ which \textit{will} be observed again by $\tilde{B}_t$ we can thus write
\begin{align}
\tilde{B}_t \ \mid \ \tilde{X}_t & \sim \text{Bin}(\tilde{X}_t, \beta_Y/(1 - \gamma_Y))\\
A_t^{(j)} & \stackrel{\text{iid}}{\sim} \text{Geom}(1 - \gamma_Y),\ \ \ j = 1, \dots, \tilde{B}_t\\
\tilde{V}_{t + i, i} & = \sum_{j = 1}^{\tilde{B}_t} I(A_t^{(j)} = i),\ \ \ i = 1, \dots
\end{align}
Combining this with equations \eqref{eq:X_tilde}, \eqref{eq:V} and \eqref{eq:U_tilde} to
\begin{equation}
\tilde{X}_t = \sum_{i = 1}^\infty \tilde{V}_{t, i} \ \ + \ \ \tilde{U}_{t, \cdot}\label{eq:reform_Xtilde}
\end{equation}
one can see that $\{\tilde{X}_t\}$ indeed follows the form \eqref{eq:inar_inf}--\eqref{eq:coupling_thinning_geom} with parameters $\lambda_Y, \beta_Y, \gamma_Y$ (the $\tilde{V}_{t, i}$ in equation \eqref{eq:reform_Xtilde} correspond to the $\alpha_{Y, i} \circ X_{t - i}$ from equation \eqref{eq:inar_inf} and $\tilde{U}_{t\cdot}$ corresponds to $W_t$). Also it is clear that both $\tilde{U}_{t, \cdot}$ and $\tilde{V}_{t + i, i} \mid \tilde{X}_t$ are independent of the past history of the observed process $\{\tilde{X}_t\}$, i.e. $\tilde{X}_{t - k}$ and $\alpha_j \circ \tilde{X}_{t - k}, i, j, k = 1, 2, \dots$

Statement (B) follows directly as it is easily verified using (A) that the given $\{\tilde{Y}_t\}$ and $\{X_t\}$ are equivalent. The restriction $0 < \beta_X < 1 - \gamma_X < 1$ ensures that $\lambda_Y > 0$ and $0 < \alpha_Y, q_Y \leq 1$ so that $\{Y_t\}$ and $\{\tilde{Y}_t\}$ are well-defined.

Statement (C) follows from statements (A) and (B) in the following way. Consider an INAR($\infty$) process $\{X_t\}$ and $\{\tilde{X}_t = q_X \circ X_t\}$. Using (B) an INAR(1) process $\{Z_t\}$ with parameters $\lambda_Z = \lambda_X(\beta_X + \gamma_X)(1 - \gamma_X)/\beta_X, \alpha_Z = \beta_X + \gamma_X$ can be constructed so that $\{\tilde{Z}_t = q_Z \circ Z_t\}; q_Z = \beta_X/(\beta_X + \gamma_X)$ is equivalent to $\{X_t\}$. Thus $\{\tilde{X}_t\}$ is in turn equivalent to $\{Z^* = (q_Zq_X) \circ Z_t\}$. One can now choose any $q_Y \in [q_Xq_Z, 1]$ and define $\{Y_t = (q_Xq_Z/q_Y) \circ Z_t\}, \{\tilde{Y}_t = q_Y \circ Y_t\}$. The process $\{\tilde{Y}_t\}$ is then obviously equivalent to $\{Z^*_t\}$ and thus $\{\tilde{X}_t\}$. The proof is complete as statement (A) implies that $\{Y_t\}$ has a representation as an INAR($\infty$) process with parameters
\begin{align*}
\lambda_Y & = \frac{\lambda_Z \frac{q_Xq_Z}{q_Y}}{1 - \alpha_Z\left(1 - \frac{q_Xq_Z}{q_Y}\right)} = \frac{\frac{\lambda_X(\beta_X + \gamma_X)(1 - \gamma_X)}{\beta_X}\cdot \frac{q_X}{q_Y}\cdot \frac{\beta_X}{\beta_X + \gamma_X}}{1 - (\beta_X + \gamma_X)\left(1 - \frac{q_X}{q_Y}\cdot \frac{\beta_X}{\beta_X + \gamma_X}\right)} =  \frac{\lambda_X(1 - \gamma_X) \frac{q_X}{q_Y}}{1 - \gamma_X - \left(1 - \frac{q_X}{q_Y}\right)\beta_X}\\
\beta_Y & = \alpha_Z\frac{q_Xq_Z}{q_Y} = (\beta_X + \gamma_X)\frac{q_X\frac{\beta_X}{\beta_X + \gamma_X}}{q_Y} = \beta_X\frac{q_X}{q_Y}\\
\gamma_Y & = \alpha_Z\left(1 - \frac{q_Xq_Z}{q_Y}\right) = (\beta_X + \gamma_X)\left(1 - \frac{q_X\frac{\beta_X}{\beta_X + \gamma_X}}{q_Y}\right) = \beta_X + \gamma_X - \beta_X\frac{q_X}{q_Y} = \gamma_X + \left(1 - \frac{q_X}{q_Y}\right) \beta_X.
\end{align*}

\section*{Acknlowledgements}
The author would like to thank Leonhard Held, Ma{\l}gorzata Roos and Christian H. Wei{\ss} for helpful comments.

\end{document}